# Remote detection of radioactive material using a short pulse $CO_2$ laser

A. Zingale[1,2], S. Waczynski[1] I. Pogorelsky[3], M. Polyanskiy[3], J. Sears[4], R. E. Lakis[2], H. M. Milchberg[1,*]

[1]*Institute for Research in Electronics and Applied Physics, University of Maryland, College Park, Maryland 20742, USA*
[2]*Los Alamos National Laboratory, Los Alamos, New Mexico 87545, USA*
[3]*Accelerator Test Facility, Brookhaven National Laboratory, Bldg. 820M, Upton, NY 11973, USA*
[4]*Lawrence Livermore National Laboratory, Livermore, California 94550, USA*
\* *milch@umd.edu*

**Abstract:** Detection of radioactive material at distances greater than the radiated particle range is an important goal with applications in areas such as national defense and disaster response. Here we demonstrate avalanche-breakdown-based remote detection of a 3.6 mCi α-particle source at a stand-off distance of 10 m, using 70 ps, long wave infrared (λ=9.2 µm) $CO_2$ laser pulses. This is ~10× longer than our previous results using a mid-IR laser. The primary detection method is direct backscatter from microplasmas generated in the laser focal volume. The backscatter signal is amplified as it propagates back through the $CO_2$ laser chain, enhancing sensitivity by >100×. We also characterize breakdown plasmas with fluorescence imaging, and present a simple model to estimate backscattered signal as a function of the seed density profile in the laser focal volume. All of this was achieved with a relatively long drive laser focal geometry ($f/200$) that is readily scalable to > 100 m.

## I. Introduction

Long range stand-off detection of radioactive material is a goal of interest to a number of communities including defense and disaster response. Modern detectors of ionizing radiation rely on the radioactive decay particle or photon interacting with the detector directly to produce the signal. However, geometric fall-off of the flux with distance, plus scattering and absorption losses, limit the practical stand-off distance of these detectors to a few tens of meters. Consequently, there has been increasing interest in developing new tools for the remote detection of radioactive material [1–6]. Recently, our group has demonstrated a new detection method that relies on laser-driven electron avalanche and the increased atmospheric ionization near a radioactive source [4,7]. As the decay particles or photons propagate away from the source they inevitably ionize air molecules, producing free electrons and negative ions. Our method uses those electrons and negative ions as seeds of laser-driven electron avalanche of air. The initial seed density in the focal volume influences the avalanche plasma dynamics and the scattered laser light. Analysis of the scattered light enables detection of radioactive material near the laser focal volume.

The choice of a long wavelength laser has two key advantages. First, the electron heating rate at intensity $I$ scales as $I\lambda^2$ [8], so that avalanches are driven more efficiently with longer wavelengths. Second, because the method must detect electrons and negative ions at ultralow concentrations of 1 part in $10^{16}$, even a very low ionization yield from neutral air directly from the laser itself could seed avalanches and swamp the desired backscatter signal. The low photon energy of long wave IR lasers guarantees much lower ionization probabilities than with near-IR lasers. As an example, at 1 TW/cm$^2$, the probability of multiphoton ionization of oxygen at λ=9.2 μm is $\sim10^{-23} \times$ that at $\lambda = 1$ μm. While our prior avalanche detection results [4,7] used a mid-IR laser (λ=3.9 μm), realizing the two advantages associated with longer wavelength, current state-of-the-art mid-infrared (mid-IR) sources do not have the pulse energy necessary to drive avalanche at distances $> \sim 10$ m [9]. The natural choice of laser source becomes $CO_2$ systems due to the availability of joule-level energy pulses in the picosecond pulse duration range, which can reasonably reach the avalanche threshold at km-scale range [10]. And the wavelength scaling advantages are further improved at λ ~10 μm.

Here, we report on radiation detection using 100 GW-scale $CO_2$ laser pulses at a stand-off distance of 10 m, a range limited by the size of the laser room. Even so, this range is 10 × longer than in our earlier work [4,7]. The pulse duration of 70 ps limits avalanche plasma growth to ~10 μm around each seed electron, leading to the generation of large numbers of plasma balls in the focal volume that scatter the incoming laser when the plasma density reaches a fraction of the critical density ($\sim 10^{19}$ cm$^{-3}$) for $\lambda = 9.2$ μm light. The direct backscatter from these micro-plasmas was amplified as it propagated back through the laser and detected 48 m upstream without any additional collection optics. This signal was analyzed for detection of the radioactive source. We present a simple model for the laser backscatter from large numbers of plasmas and relate the signal to the seed density in the focal volume. Additionally, side-on images were taken that show the visible fluorescence brightness scales with the initial ion seed density. These results were achieved with a relatively long drive laser focal geometry ($f/200$) that is readily scalable to > 100 m.

## II. Experimental Setup

Figure 1 shows a schematic of the experimental setup. The laser used to drive avalanche for this experiment was the $\lambda = 9.2$ μm, picosecond $CO_2$ laser at the Accelerator Test Facility (ATF) at Brookhaven National Laboratory (BNL) [11]. For the purpose of this experiment, the laser was configured to output an uncompressed pulse of ~70 ps duration. The 3.6 mCi Po-210 test source emitted 5.3 MeV alpha particles. The laser was focused at $f/200$ at 10 m distance, giving a 2.1 mm full width at half maximum (FWHM) diameter focal spot and a Rayleigh range of 1.1 m. The energy per pulse was varied in the range 0.1 to 3 J, corresponding to peak intensity in the range 0.05 – 1.2 TW/cm$^2$. We define the avalanche threshold intensity, $I_{th}$, as the intensity for which the heating rate exceeds the loss rate (for plasma cooling and particle diffusion) for a sufficient duration that the breakdown plasma is visible by eye. Because the heating rate scales as $I\lambda^2$, $I_{th} \propto \lambda^{-2}$. In our earlier work [7] with $\lambda = 3.9\ \mu m$, we found $I_{th} = 1$ TW/cm$^2$ so that at $\lambda = 9.2$ μm, $I_{th}\sim 0.2$ TW/cm$^2$ [4], assuming similar loss mechanisms. The avalanche plasma region occupies the volume $V_{th}$ where $I > I_{th}$, where $V_{th}\sim 6$ cm$^3$, with length ~2 m and diameter ~2 mm. The plasma growth around a single avalanche seed electron is limited by diffusion to a radius of < 10 μm for the 70 ps pulses and laser intensity of this experiment [7,12]. Therefore, the plasma region will be made up of a collection of micro-plasmas centered on each avalanche seed (shown in Fig. 2(c)), unless the average spacing between seeds is smaller than the diffusion length during plasma growth. For a 10 μm diffusion length this corresponds to a seed density $\rho_s \sim 10^9$ cm$^{-3}$. The Po-210 source produces a negative ion density ~10$^6$ cm$^{-3}$ within its stopping distance in air (< ~3 cm). The background seed density, which is generated from cosmic rays and other environmental radiation sources like radon gas, is ~10$^3$ – 10$^4$ cm$^{-3}$ [7].

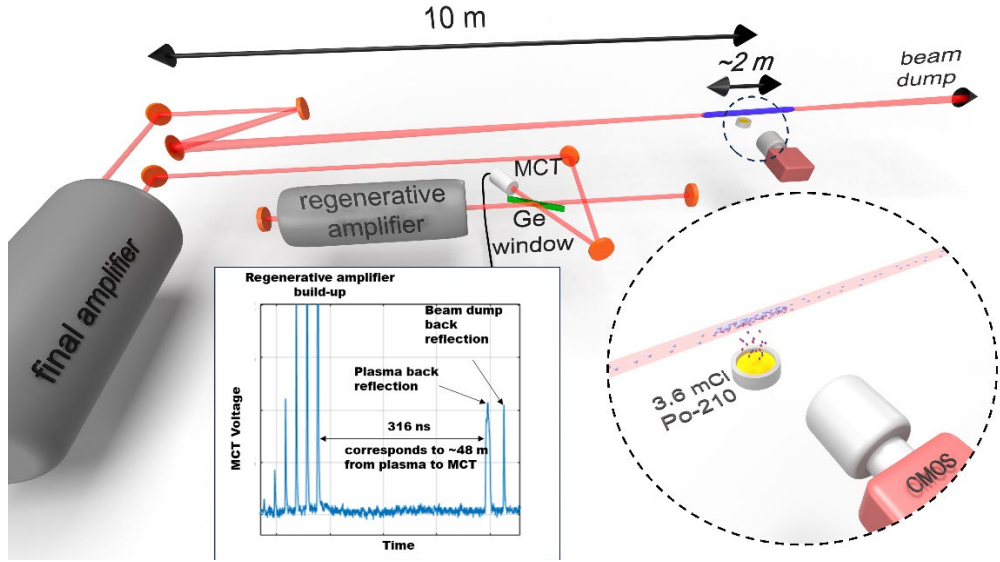

**Figure 1.** Experimental setup. A ~100 picosecond FWHM, <5 J, λ=9.2 μm laser pulse is focused from a spherical mirror at ~$f/200$, producing a 2.1 mm FWHM focal spot 10 m from the focusing optic. The confocal parameter of the beam was ~2 m. A 3.6 mCi, Po-210 source was placed at various longitudinal and transverse distances from the beam waist. The directly backscattered light was collected with the focusing optic and measured with a ~1 ns risetime MCT photo-detector in the regenerative amplifier (RGA). The Ge window is a near-IR laser-activated photoconductive switch for ejecting pulses from the RGA A trace from this detector is shown with features produced from the laser amplification process (build-up), reflection from the plasma at focus, and reflection from the beam dump. The time-delay between the build-up and plasma reflection corresponds to ~48 m optical path-length separation between the focus and the detector. There were no collecting optics used at the detector, only a bare 1 mm$^2$ active area MCT sampling the ~25 mm beam. A CMOS camera was also used to directly image the plasma fluorescence with a 38 mm diameter collection lens. It was either ~200 mm or ~1000 mm from the plasma region as indicated in the text.

There were two primary diagnostics of the avalanche plasmas: a CMOS camera for direct imaging of the plasma fluorescence, and a mercury cadmium telluride (MCT) photodiode in the laser chain normally used to monitor amplification in the regenerative amplifier. In this case, the MCT also measured relative intensity of direct backscatter. The CMOS camera was operated at two transverse distances (~20 cm and ~1 m) from the plasma for high- and low-resolution images. In the low-resolution imaging, individual microplasmas cannot be distinguished as in the high-resolution case.

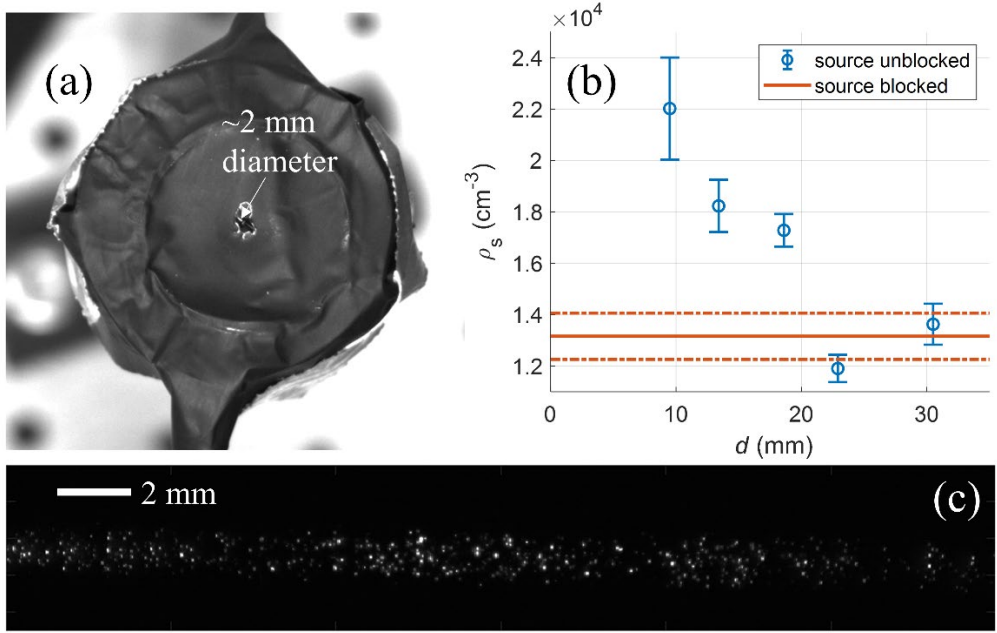

**Figure 2.** Seed density measurement via breakdown counting. (a) Image of Po-210 blocked with an aluminum foil with a ~2 mm hole, making the source point-like and reducing the activity by ~100× to ~30 µCi. (b) Plot of seed density, $n_s$, as a function of the transverse distance of the source from the laser focus. The blue circles represent the measured seed density averaged over 0 shots. The error bars represent the standard error. The solid orange line represents the measured background density (source blocked completely) with the dashed lines showing the standard error. (c) Image of breakdown plasmas in the section of the focal volume directly above the source.

## III. Experimental results

### A. Plasma fluorescence images

The Po-210 source produces seed densities high enough that it is difficult to distinguish individual breakdown sites in the images, even though they may be well separated in space. This is due to the projection of the 3D distribution of plasmas onto the 2D sensor surface. In order to generate a countable number of breakdowns, the effective activity was reduced by about 100× (~30 µCi) by covering the source with an aluminum foil with a ~2 mm diameter hole as shown in Fig. 2(a). This enables a precise seed density measurement as a function of source transverse distance from the laser focal volume. The volume containing avalanche microplasmas was taken as a cylinder of length equal to the width of the image and a diameter equal to the distance from the highest located to the lowest located breakdown in the image. The laser pulse energy ranged from 1 to 2.5 J in this case.

A custom particle counting algorithm was used to count individual breakdowns (see Supplemental Material [13]). The peaks associated with each breakdown site were isolated using a technique called "morphological opening" with a circular structure element [14] with a ~100 µm diameter. This effectively removes all features larger than 100 µm (much larger than the expected breakdown size). Then a single breakdown was defined as a contiguous region with pixel count >5× the mean pixel value in regions where no breakdowns are present. Figure 2(b) plots the number of breakdown sites per unit volume $\rho_s$ (equivalent to initial seed electron density) vs. source transverse distance $d$ from the laser focus. Each point is the average of 10 shots. The background seed density was measured as $1.3 \times 10^4$ cm$^{-3}$ with the Po-210 source blocked. The density increases as $d$ decreases, as expected, approaching background density for $d > 20$ mm. The observed stopping distance is less than the expected ~3 cm because of energy loss in a gold flash-plate deposited on top of the Po-210 layer to prevent contamination. No significant Bragg peak is observed in the density profile because the geometric fall-off of the α-particle flux with $d^2$ is a stronger effect than the increase in energy deposition per unit length characterized by the Bragg peak. The maximum measured density of breakdown sites is about $2.2 \times 10^4$ cm$^{-3}$, only 1.7× the background seed density and $10^{-15}$ of atmospheric density, but still distinguishable because each individual breakdown site corresponds to one seed charge. That is, each avalanche breakdown begins with a single free electron liberated from a negative ion early in the laser pulse. While direct imaging of the focal volume cannot be scaled to long stand-off distances due to the unattainable resolution required, this measurement confirms that the number of breakdown sites in the laser focal volume scales with the number of source-induced seeds.

As long as the seed ion density is well below the ~$10^9$ cm$^{-3}$ estimate for breakdown overlap discussed above, we can reasonably treat each breakdown site as independent of the others, and the fluorescence intensity emitted from each breakdown site will depend primarily on drive laser intensity. For singly ionized air constituents, the plasma density is ~$5 \times 10^{19}$ cm$^{-3}$, ~100× below the critical plasma density for the optical wavelength photons emitted by the plasma. Therefore, neighboring breakdowns will not significantly attenuate the optical wavelength emission. This suggests that even for low resolution imaging, which would be necessary at long range, the integrated brightness will scale linearly with the total number of seed ions, $N_s = \rho_s V_{th}$, in the focal volume.

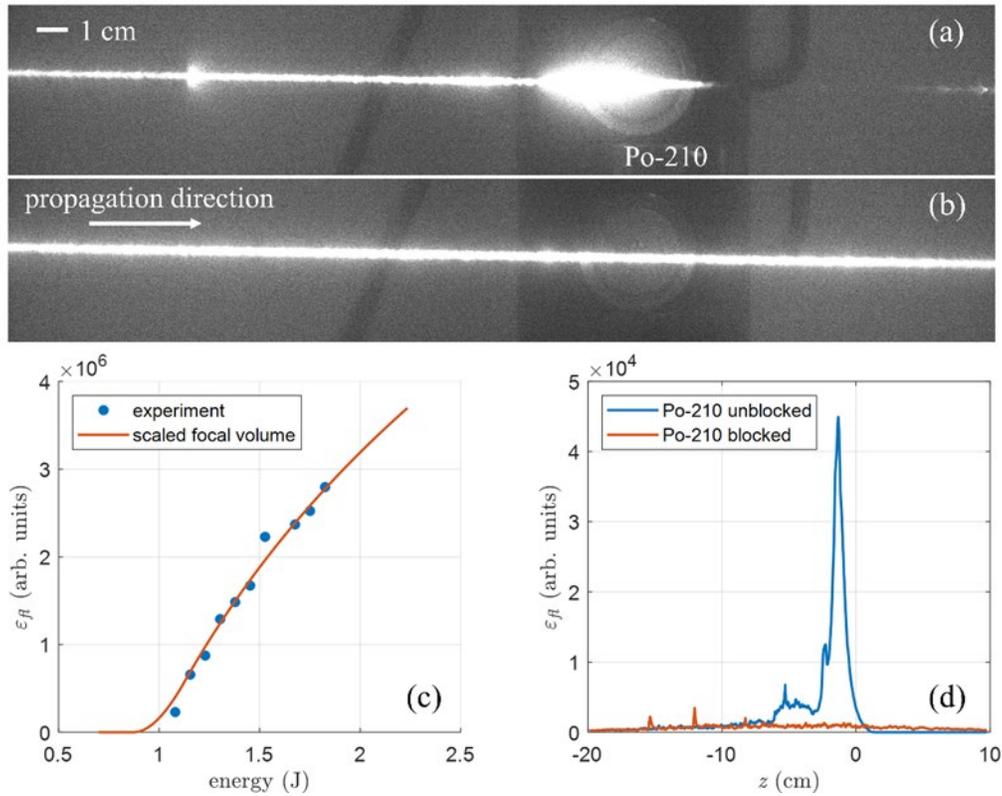

**Figure 3.** Seed density measurement via integrated brightness of visible plasma fluorescence. (a) Side-on image of breakdown fluorescence with the Po-210 source $d = 25$ mm from laser focus. (b) Similar image with the Po-210 source blocked, where breakdowns are seeded by background ions and electrons. In both (a) and (b) the width of the field of view is about 30 cm and no saturated pixels were observed. (c) Plot of integrated fluorescence intensity, $\mathcal{E}_{fl}$, (blue circles) vs. drive pulse energy. We also show the scaling of $V_{th}$ with energy (orange line) calculated for a Gaussian beam. (d) Sample single shot plot of integrated fluorescence intensity, $\mathcal{E}_{fl}$, vs. position along the laser propagation axis, $z$, for Po-210 exposed (blue) and blocked (orange).

To demonstrate this scaling, we imaged the plasma from ~1 m away and removed the pinhole mask on the Po-210 source. Based on the solid angle subtended by the collection lens and assuming isotropic emission, we expect a light collection efficiency ~$10^{-4}$, and the resolution of 75 μm/pixel did not allow discrimination of individual breakdowns because each breakdown radius < ~$10\ \mu m$. Typical images are shown in Fig. 3(a) and (b) with the source exposed and blocked respectively. It is evident that the plasma emission brightness above the source is much higher than regions away from the source. Above the source, the density of breakdown sites is high enough to significantly attenuate the $CO_2$ laser beam (propagating left to right) and greatly reduce the brightness of downstream breakdowns relative to the upstream region. Figure 3(c) plots integrated brightness for shots with the source blocked as a function of input laser pulse energy. For increased laser energy, $V_{th}$ also increases. The intensity profile of a Gaussian beam, $I(r,z) = I_0(w_0/w(z))^2 \exp[-2r^2/w(z)^2]$, was set equal to the avalanche intensity threshold, $I_{th}$, and then solved for the effective threshold radius $r_{th}(z)$, from which the threshold volume $V_{th}$ can be computed. The integration was performed numerically over the range of $z$ in the images. The result was then scaled to the experimental value at 1.4 J and plotted with the orange line. The total fluorescence energy is $\mathcal{E}_{fl} = \rho_s V_{th} \bar{\varepsilon}_b$, where $\bar{\varepsilon}_b$ is the average fluorescence energy from an individual breakdown site. The good agreement in Fig. 3(c) shows that the total brightness scales with $V_{th}$ as expected, because these are shots with the source blocked and therefore have uniform background seed density and $\bar{\varepsilon}_b$ is not a strong function of pulse energy.

Now we use the integrated fluorescence brightness along with the background seed density measurement shown in Fig. 2(b) to estimate the seed density in the region near the source. Figure 3(d) plots integrated fluorescence intensity, $\mathcal{E}_{fl}$, vs. position along the laser propagation axis, $z$. These curves were obtained by taking the images in Fig. 3(a) and (b), subtracting the noise level, and then integrating over the vertical dimension. This permits a comparison of the peak $\mathcal{E}_{fl}$ between Po-210 source -unblocked and -blocked shots. The peak value for $\mathcal{E}_{fl}$ in the unblocked case is ~75× greater than that of the blocked case. Scaling the background seed density of $1.3 \times 10^4$ cm$^{-3}$ by 75× gives an estimate of the peak seed density above the source as $\rho_s \sim 9.8 \times 10^5$ cm$^{-3}$. This is in good agreement with previous observations [4,7] using the same source at 25 mm transverse distance. It is notable that the peak breakdown fluorescence occurs a few cm upstream of the center of the source ($z = 0$) even though the peak seed density is greatest at $z = 0$. This is because significant scattering by upstream avalanche microplasmas reduces the downstream laser intensity below $I_{th}$, as can be seen in Fig 3(a). This implies that ~$10^6$ cm$^{-3}$ is a rough threshold for maximum measurable seed density; avalanche microplasmas from higher seed densities will effectively prevent the laser from exciting avalanches throughout $V_{th}$.

## B. Laser backscatter detection

The critical plasma density is $N_{cr} = 1.3 \times 10^{19}\ cm^{-3}$ at $\lambda = 9.2\ \mu m$. Our avalanche simulations described in Refs. [4,15], and applied to the laser conditions in this experiment (see Supplement 1), estimate that the avalanche plasma peak temperature is ~20 eV, exceeding the first ionization energy of air constituents, so we assume saturation of single ionization corresponding to the number density of air at STP, $N_{air} = 5.4 \times 10^{19}\ cm^{-3} > N_{cr}$. Therefore, all breakdowns strongly scatter the incoming laser as they approach and exceed the critical plasma density.

An optimal diagnostic for remote detection is one that is always aligned to observe the region of interest, in this case the laser focal volume. Light scattered back toward the focusing element has this feature, and it can propagate backward through the laser amplifier. In this experiment, we estimate >100× residual gain in the linear regime (see Supplement 1). The amplified backscatter then propagates upstream and transmits through the Ge window in the RGA, and is detected by the MCT (see Fig. 1). Note that the photoconductive plasma in the Ge window, which acts as an optical switch, decays over ~1 ns; the window will be fully transparent when the backscattered light arrives after >300 ns [11,16].

Analysis of the MCT traces provides information about the microplasmas in the focal region. The MCT trace in Fig. 1 is typical: the early spikes (left side) are reflections from the Ge window and show pulse amplification buildup in the RGA. The next spike, 316 ns later, is backscatter from the microplasmas in the focal volume, and the final spike is backscatter from a beam dump positioned ~10 m downstream of the focus. The RGA build-up and beam dump spikes were ~1.5 ns FWHM, limited by the nanosecond temporal response of the MCT to a 70 ps signal. However, the plasma backscatter spike was typically ~12 ns FWHM for shots with the source blocked, significantly lengthened by the axially distributed backscattering, and giving an estimated plasma length of ~1.8 m. This is consistent with the earlier ~2 m plasma length estimate based on Gaussian beam geometry.

To discern the presence of the radioactive source, we measure the total amplified backscattered signal, $E_{bs}$, by integrating the MCT trace in a 30 ns temporal window centered on the plasma backscatter spike. Figure 4(a) plots $E_{bs}$ vs. the longitudinal position of the source, as illustrated by Fig. 4(c). Here, we redefine $z = 0$ to be the earliest upstream axial location in the laser beam where plasma appears in a shot with the alpha source blocked. The transverse distance of the alpha source from the laser beam axis was always $d = 2.5$ cm. The absolute backscattered signal depends on the drive pulse energy, which was monitored on each shot. The data in Fig. 4(a) represents an energy range of 0.5 to 2.5 J. The effect of this variation on the signal was subtracted as follows: a linear fit was performed on $E_{bs}$ vs. pulse energy for all shots with the source blocked, then that fit was used to subtract the expected background contribution to $E_{bs}$ for all shots. This procedure gives an average signal of zero for the shots with the alpha source blocked.

It is seen that increasingly downstream placement of the source gives higher total backscattered energy, with a significantly higher signal for >80 cm. This trend arises because the region near the source significantly attenuates the laser, reducing plasma generation and backscatter from the downstream region, as described earlier. Moving the source downstream enables contribution of increasing upstream background to the total signal. The total backscatter signal is the sum of signals from the background and alpha source regions (Fig. 4(c)). If the source is positioned near $z = 0$ cm (the upstream edge of the focal region), it nearly eliminates the background region scatter. When the source is located at the downstream edge near $z = 200$ cm, the full background signal contributes to the total signal. However, the maximum accessible position in the experiment was $z = 120$ cm due to space constraints in the laser room.

The total backscattered signal above background at >80 cm is what enables detection of the alpha source at a stand-off distance of 10 m. We note that the low signal-to-noise ratio is from the small length of the signal region compared to the background region. For $d = 2.5$ cm, the alpha source generates an elevated seed density over a ~3 cm length, which is only 1.5% of the ~2 m confocal parameter, showing that the signal backscattering per unit length is approximately two orders of magnitude greater than the background.

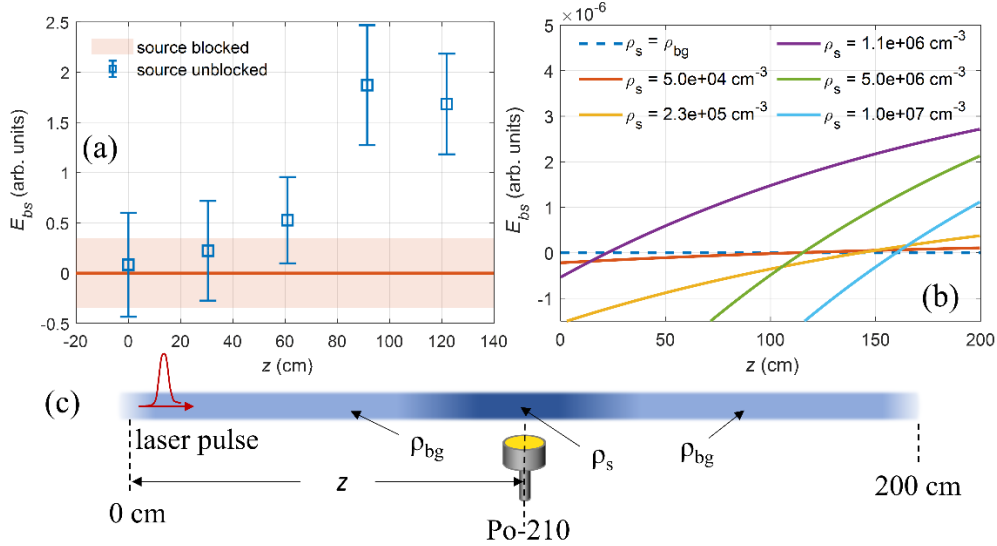

**Figure 4.** Backscattered energy, $E_{bs}$, as a function source position within the focal volume, $z$. (a) A scatter plot of $E_{bs}$ vs. $z$ is shown. The mean over 10 shots at each z-position with the source unblocked are indicated by the blue squares, and the error bars represent the standard deviation of the mean. The orange line and surrounding shaded region represent the mean with the source blocked and corresponding standard deviation, respectively. (b) Simulated integrated backscatter for varying seed density above the source $\rho_s$, with background seed density $\rho_{bg} = 1.1 \times 10^4$ cm$^{-3}$. (c) Schematic layout of the coordinate system and laser pulse propagation direction with regions of background and signal seed density pointed out.

To corroborate the experimental results, we also plot the results of $E_{bs}$ calculated by the scattering model described in the following section. We assume a microplasma density of $1.1 \times 10^4$ cm$^{-3}$ for the background and a 3-cm-long signal region with microplasma densities from the background level to $10^7$ cm$^{-3}$. The trend seen in the experiment is qualitatively reproduced for all signal densities, with the best match at $1.1 \times 10^6$ cm$^{-3}$, which matches the measured density from the fluorescence imaging technique. However, the quantitative match should be interpreted cautiously, as discussed below.

## IV. Scattering model

### A. Simulation of experimental configuration

It is not straightforward to calculate the scattering cross section of individual breakdown sites because the breakdown radius ($< \sim 10\ \mu m$) is comparable to the laser wavelength of 9.2 μm. This necessitates use of Mie theory [16]. For these calculations, we use CELES, a freely available software package [17], which is an implementation of the generalized multiparticle Mie method, or T-matrix method [16]. The code computes the electromagnetic transport properties of large ensembles of spherical particles of arbitrary size, position, and index of refraction. Breakdown sites were given the same local plasma density and refractive index with variable size and randomized position. In reality, all these parameters are dynamic with significant variation on the timescale of the pulse duration. However, such a simplified model can provide physical insight into our measurements.

The size, temperature, and electron-neutral collision frequency in avalanching microplasmas were calculated with the breakdown simulations (see Supplement 1). Here, the temperature peaks at ~20 eV, so that single ionization of air ($\sim 5 \times 10^{19}$ cm$^{-3}$) is a reasonable assumption as it is between the first and second ionization energies of the dominant air constituents. The complex index of refraction was calculated with this temperature and plasma density according to the Drude model [18]. The breakdown sizes are not identical due to the local intensity variations in the beam. The breakdown radii were normally distributed with a mean of 7 μm and a standard deviation of 1 μm, based on breakdown simulations [4,15] using experimental laser parameters (see [13]).

With the above scatterer characteristics, CELES computed the forward scattered and backscattered energy fraction (or transmittance and reflectance) from a 2 mm diameter, 2 mm long cylindrical section with varying numbers of scatterers. The incident beam was a 2 mm FWHM Gaussian intensity profile. Forward and backscattered signals were defined to be the fractions of the incident beam energy scattered into the divergence angle of the beam in the forward and backward directions. CELES calculations were limited to 2 mm sections because the entire 2 m long focal volume was too computationally expensive to simulate at once.

In order to model the total backscatter from the focal volume we divided the 2 m focal volume into one-thousand 2 mm long sections, each with a fractional energy transmittance and reflectance, $\tau_n$ and $r_n$, corresponding to the seed density in $n^{th}$ section of the focal volume. Most of the focal volume (985 sections) had background density, and only 15 sections directly above the alpha source had elevated seed density. Using the values from a single 2 mm section of a given density, we can calculate the total transmittance and reflectance of a subsection of the focal volume with that density (i.e. if the upstream background region is 200 mm long, we model it as 100 identical sections with background seed density). The total reflectance $R$ of a region consisting of $N$ identical 2 mm sections is calculated by adding the reflection contributions while accounting for the transmission through all the sections upstream of each, yielding

$$R = \sum_{n=1}^{N} \tau^{n-1} r \tau^{n-1} = \frac{r(1-\tau^{2N})}{1-\tau^2}. \tag{1}$$

We ignore multiple reflections, because even for the highest seed densities $r_n \sim 10^{-5}$.

Using Eq. (1) to evaluate the upstream, source, and downstream contributions to the backscattering ($R_u, R_s, R_d$) (see Fig. 4(c)), the total reflectance or backscatter is

$$R_{total} = R_u + R_s \tau_b^{2u} + R_d \tau_b^{2u} \tau_s^{2s}, \tag{2}$$

where $\tau_b$ and $\tau_s$ are the transmittances of single 2 mm sections with background and signal seed densities, and $u$ and $s$ are the number of 2 mm sections in the upstream and source regions. For our experimental conditions, $s = 15$ (3 cm region of elevated seed density) and $u$ varies depending on source location. All the $\tau$ and $r$ values at various densities are taken from the CELES calculations, and together with equations 1 and 2 produce the plot in Fig. 4(b). The source location was varied by changing the number of upstream and downstream background sections, keeping $s$ constant.

## B. Analytical model of backscattering for constant seed density

The model represented by Eqs. (1) and (2) is suitable for our experimental situation, where the radioactive source-seeded-volume is small compared to the volume of the laser focus. A more realistic remote detection scenario would involve a source-seeded volume that is large compared to the laser focal volume, as one would expect from more spatially dispersed sources or from source particles with long stopping distances, such as γ-ray photons. For this reason, it is instructive to consider the backscattered signal fraction for a uniform seed density $\rho_s$ in the focal volume.

Consider an infinitesimal slice of width $dz$ at axial position $z$ of the scattering volume, where $z = 0$ and $z = L$ are the axial extremes of the backscattering along the laser axis, where $L$ is taken to be the laser confocal parameter. The backscattered signal contribution from this slice is $dR(z) = \tau^2(z)dr(z)$, where $dr(z) = \overline{\sigma_{bs}}\rho_s(z)dz$ is the differential reflectance from the slice at $z$, $\overline{\sigma_{bs}}$ is the backscatter cross section averaged over the microplasma size distribution, and $\tau(z) = \exp[-\int_0^z \mu(z')dz']$ is the transmittance from $z = 0$ to the slice at $z$ (and from the slice back to $z = 0$) [16]. The optical density (extinction per unit length) is $\mu(z) = \overline{\sigma_{ext}}\rho_s(z)$, where $\overline{\sigma_{ext}}$ is the average extinction cross-section of the scatterers. Integrating over the confocal parameter gives the total backscatter reflectance

$$R = \int_0^L e^{-2\int_0^z \overline{\sigma_{ext}}\rho_s(z')dz'} \overline{\sigma_{bs}}\rho_s(z)dz. \tag{3}$$

For constant $\rho_s$, this yields

$$R = \frac{\overline{\sigma_{bs}}}{2\overline{\sigma_{ext}}}[1 - e^{-2\overline{\sigma_{ext}}\rho_s L}]. \tag{4}$$

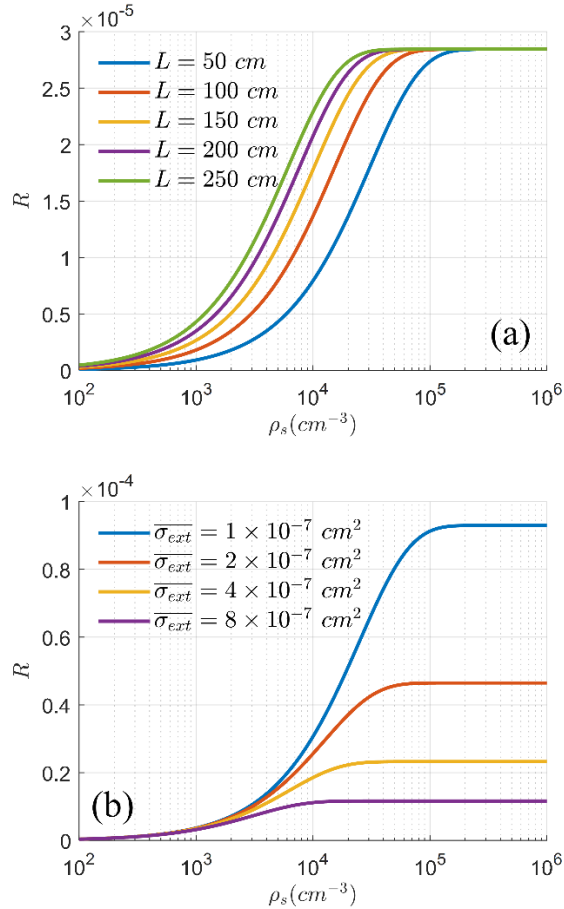

**Figure 5.** (a) Backscattered signal from Eq. (4) vs. seed density $\rho_s$ and confocal parameter $L$ for $\overline{\sigma_{ext}} = 3.3 \times 10^{-7}$ $cm^2$. (b) Backscattered signal vs. seed density for various extinction cross-sections with $L = 200$ $cm$.

CELES calculates the total far field power distribution of a section of finite length ($\Delta z = 2\ mm$) directly by expanding the field in terms of spherical vector wave functions and solving for the expansion coefficients. Over a finite length section $\Delta z$ of constant $\rho_s$, $dr(z)$ and $\tau(z)$ simplify to $\Delta r = \overline{\sigma_{bs}}\rho_s\Delta z$ and $\tau = \exp[-\mu\Delta z]$. Using the definitions of forward- and backscattered signal fractions from the previous section, we can integrate this distribution over the solid angle of the beam in the forward and backward directions to get $\Delta r$ and $\tau$. Using these values, we simply rearrange to obtain the cross sections as $\overline{\sigma_{bs}} = \Delta r/(\Delta z\rho_s) = 3.4 \times 10^{-12}\ cm^2$ and $\overline{\sigma_{ext}} = -\log(\tau)/(\Delta z\rho_s) = 2.4 \times 10^{-7}\ cm^2$. These cross sections are properties of the breakdown plasmas and therefore will only vary with $\rho_s$ if multiple scattering is significant. We found that this was not the case: the cross sections did not vary significantly over the range of $\rho_s$ simulated here.

Figure 5 plots Eq. (4) for various values of $L$ and $\overline{\sigma_{ext}}$. It is evident from Eq. (4) that the increase of $R$ with $\rho_s$ will start to saturate when $\rho_s > \sim 1/(2\overline{\sigma_{ext}}L)$. In that case, upstream extinction is strong enough to significantly reduce the downstream contribution to the total backscatter signal. For $\rho_s < 1/(2\overline{\sigma_{ext}}L)$, the total backscatter signal increases strongly with $\rho_s$. As shown in Fig. 5(a), the backscatter signal saturates near $\rho_s > 10^5\ cm^{-3}$ for an $L = 50$ cm long focal volume, which corresponds to $\sim f/100$ focal geometry for a Gaussian beam. This is a higher saturation point compared to a longer focal volume, meaning the overall dynamic range of this technique improves for shorter focal volumes. Similarly, smaller $\overline{\sigma_{ext}}$ increases the saturation density, as seen in Fig. 5(b). $\overline{\sigma_{ext}}$ is a property of the microplasmas that will increase with plasma density and size owing to greater scattering and absorption. It is possible that these plasma properties can be influenced by laser pulse parameters. For instance, using a shorter pulse duration could lead to less plasma growth, smaller breakdowns and therefore smaller $\sigma_{ext}$, again increasing the dynamic range of the technique. This is true independent of the backscatter cross section. The saturation density only depends on the extinction cross-section and length of the focal volume.

## V. Conclusions

We have demonstrated detection of a 3.6 mCi Po-210 alpha particle source at a stand-off distance of 10 m using a ~1 J, 70 ps, λ=9.2 μm laser pulse. The pulse duration was short enough to limit plasma diffusion to ~10 μm, causing individual seed electrons and ions to produce individual microplasmas. Laser light backscattered into the laser system was amplified by residual energy in the gain medium before being detected by an MCT photodiode, enhancing detection sensitivity by greater than 100 ×. Measurements from the MCT are consistent with full Mie calculations of the laser scattering of large ensembles of spherical plasmas of varying size. We also verified that the total plasma fluorescence scales with seed density, so resolution of the individual plasma sites is not necessary to measure the seed density from long standoff distances.

The results presented here provide insight into what is required to extend avalanche-based radiation detection to γ-ray sources and longer stand-off distances. First, due to the 10 meter-scale range of γ-ray photons in air, the energy deposition per unit volume and seed density is much lower. For example, a Cs-137 source of similar activity to the Po-210 source used in this work would generate $\rho_s \sim > 10^5\ cm^{-3}$ 1 meter from the source, which would cover the entire experimental confocal parameter $L \sim 200$ cm in this work and nearly saturate the backscattered signal. Keeping the focal geometry constant and extending to 100 m would simply require a 0.5 m diameter focusing optic. This would allow detection of a mCi of Cs-137 at 100 m, far outside the range of the current state-of-the art detectors.

However, as we have shown, an increase in the $f/\#$ of the focusing geometry will lead to a reduction of the saturation density for the backscattered signal owing to the increased confocal parameter. We estimate that background seed density will saturate the backscatter signal when the confocal parameter exceeds ~500 cm, corresponding to ~$f/650$. This requires a ~1.5 m diameter focusing optic and >~2.5 J pulse energy at 1 km. The scaling of the confocal parameter with $(f/\#)^2$ makes significantly longer detection difficult with the backscatter method.

In contrast, measurement of the visible fluorescence scales favorably with longer focal geometry because the fluorescence intensity scales as $V_{th} \propto (f/\#)^4$. This implies that for a given input beam diameter, longer focal lengths lead to *increased* signal collected because $\varepsilon_{fl} \propto V_{th}$ whereas the geometric fall-off of the collected signal will scale as $(f/\#)^2$. This ignores atmospheric attenuation because even on the km scale, the atmospheric extinction coefficient applied to plasma fluorescence light (at $\lambda \sim 450\ nm$) is ~0.2 km$^{-1}$ at sea level [19]. The primary hurdle is detecting the visible fluorescence above noise, especially during daylight conditions. A detailed study of spectrally resolved fluorescence measurements is therefore of great interest for this technique, and will be the subject of future work.

The authors thank staff members at Brookhaven National Lab Accelerator Test Facility, and R. Schwartz for technical assistance. This work is supported by the National Science Foundation (PHY2010511) and the National Nuclear Security Administration's Office of Defense Nuclear Nonproliferation, and the US DOE Office of Science under contract DE-SC0012704.

# Remote detection of radioactive material using a short pulse CO₂ laser: supplemental material

A. Zingale[1,2,*], S. Waczynski[2] I. Pogorelsky[3], M. Polyanskiy[3], J. Sears[4], R. E. Lakis[1], H. M. Milchberg[2]

[1]*Los Alamos National Laboratory, Los Alamos, New Mexico 87545, USA*
[2]*Institute for Research in Electronics and Applied Physics, University of Maryland, College Park, Maryland 20742, USA*
[3]*Accelerator Test Facility, Brookhaven National Laboratory, Bldg. 820M, Upton, NY 11973, USA*
[4]*Lawrence Livermore National Laboratory, Livermore, California 94550, USA*
**azingale@umd.edu*

1. **Breakdown simulations**

For the purpose of this work, we need a model to estimate the scattering properties of the breakdown sites. This requires plasma temperature, density, and size. Based on the breakdown images in this experiment as well as prior work [1,2], we expect to be in the discrete breakdown regime, where individual breakdown sites are seeded by single isolated negative ions and do not overlap or merge. We have previously developed simulations for this physical situation which are an adaptation of the 0-dimensional (0D) model presented in detail in Refs [1–3]. We briefly restate the outline of the model here, and discuss the aspects that are most relevant to this work.

The 0D model uses plasma temperature, density, laser pulse intensity, and laser wavelength to evaluate direct laser ionization, collisional ionization, electron attachment, electron-ion dissociative recombination, and collisional negative ion detachment. As described in Ref. [2], we find that the temperature follows the intensity profile of the laser nearly adiabatically with (~2 ps) equilibration time. The simulation proceeds by calculating the electron temperature $T_e$ and ionization rate $\nu_i$ at a given timestep from the laser intensity and wavelength. The total number of electrons $N$ is then grown during each timestep using the ionization rate as $\Delta N = N\nu_i \Delta t$ and $N(t + \Delta t) = N + \Delta N$ where $\Delta t$ is the timestep duration. The diffusion coefficient $D_e = k_B T_e / m_e \nu$ is also calculated, where $k_B$ is the Boltzmann constant, $m_e$ is the electron mass, and $\nu$ is the electron-neutral collision frequency. In general, the diffusion determines the average electron displacement as $r_d(t) = \sqrt{2D_e t}$. For the differential growth of the breakdown radius during the small timestep $\Delta t$, $\Delta r_d = [D_e/r_d]\Delta t$ where $r_d$ is the breakdown radius. The breakdown volume is therefore $V = \frac{4}{3}\pi r_d^3$ and the plasma density is simply $N/V$. Because diffusion is not well defined for small numbers of particles, we ignore diffusion for $N < 100$. After $N = 100$, we calculate the total time $\Delta t_{f-i}$ between $N(t_i) = 1$ and $N(t_f) = 100$. Then we begin the differential growth using the "initial" radius $r_i = \sqrt{2D_e \Delta t_{f-i}}$. The choice of $N = 100$ is arbitrary but we find the final breakdown size is insensitive to this choice.

The growth is arrested when the plasma Debye length is equal to the breakdown radius, and electrostatic forces from charge separation at the plasma edges restrict electron diffusion. The ambipolar diffusion that follows is negligible on the 70-picosecond timescale of the pulse. This model predicts a remarkable insensitivity of the breakdown radius to the peak pulse intensity for a gaussian temporal profile. This is because the breakdown process is relatively fast compared to the pulse duration. So, the breakdown occurs quickly once the intensity exceeds the breakdown threshold. This means the breakdown may happen at different parts of the intensity envelope but the breakdown still proceeds at roughly the same intensity regardless of the peak intensity of the pulse. And to reiterate, once the breakdown is fully developed, the ambipolar diffusion length over the subsequent part of the pulse is small compared to the initial electron diffusion. The model predicts breakdown radii between 6 and 8 µm for a full order of magnitude intensity range of 0.2 – 2 TW/cm$^2$, which corresponds to the experimental intensity range. A mean breakdown radius of 7 µm and standard deviation of 1 µm were therefore chosen for the scattering model calculations.

## 2. Breakdown counting algorithm

The images of the visible fluorescence of the breakdown sites allow a calculation of the initial seed density as each breakdown site marks the location of a single negative ion on average. In order to count the breakdown sites programmatically a particle counting algorithm was developed. Breakdowns are circular peaks < ~10 pixel (~100 µm) FWHM diameter in these images. The height of the feature above the immediate surrounding area (background) of the image is the relevant criteria for defining a peak. But if the background level varies significantly a simple threshold cannot accurately determine the presence of a peak. A dynamic background subtract is necessary. One way to do identify all features that we are not interested in (background) is to perform an operation called "morphological opening" [4]. We use the "imopen" function in MATLAB as our implementation. This technique replaces each pixel value with the *minimum* in its neighborhood ("erosion"). The neighborhood is defined by an arbitrary structure element (a disk with a diameter of 10 pixels in this case). Then the process is repeated but replacing each pixel with the *maximum* in its neighborhood ("dilation") the second time. The combined effect is to remove all features < 10 pixel diameter, leaving larger features intact. The resulting image is an accurate background which is subtracted from the original image leaving the peaks corresponding to each breakdown. The image is then binarized with a simple threshold value and continuous regions can be distinguished and counted with the "regionprops" function in MATLAB.

## 3. Backscatter amplification estimate

Here we analyze the backscatter amplification process to estimate the gain factor seen by the backscattered light in the experiment. The CO$_2$ laser upper state has a relaxation time of ~1 µsec [5], much longer than the ~116 ns (~35 m) round trip propagation time from the laser output to the focal volume. The final amplifier has a small signal gain of 2%/cm, a total propagation length of 500 cm, and a saturation energy of ~20 J [6]. We operated at a maximum energy of 2.5 J, well below saturation, and the round-trip time-of-flight of the pulse to the focal volume was ~10% of the characteristic relaxation time of laser transition. Therefore, we can make a conservative estimate for the residual gain assuming the density of excited-state molecules drops by half with the stimulated emission cross-section remaining constant. This corresponds to a gain coefficient of 1%/cm which leads to a gain factor of 148× in the final amplifier; a significant increase in the backscatter signal. Because the backscattered energy fraction is low (~10$^{-5}$), we expect to be well within the small signal gain regime and the detected backscattered signal is proportional to the backscattered energy from the focal volume.